\documentclass[conference, a4paper]{IEEEtran}

\usepackage{cite}
\usepackage{amsmath,amssymb,amsfonts}
\usepackage{graphicx}
\usepackage{textcomp}
\usepackage{xcolor}
\usepackage{array}
\usepackage{booktabs}
\usepackage{multirow}
\usepackage{url}
\usepackage{balance}
\usepackage{subcaption}
\usepackage{tikz}
\usepackage{pgfplots}
\pgfplotsset{compat=1.17}
\usetikzlibrary{positioning, arrows.meta, fit, backgrounds, calc, shapes.geometric, patterns, decorations.pathreplacing}
\definecolor{nfblue}{HTML}{4472C4}
\definecolor{nforange}{HTML}{ED7D31}
\definecolor{nfgreen}{HTML}{70AD47}
\definecolor{nfpurple}{HTML}{7030A0}
\definecolor{nfgray}{HTML}{A5A5A5}
\definecolor{r17red}{HTML}{C00000}

\begin{document}

\title{Serverless5GC: Private 5G Core Deployment via a Procedure-as-a-Function Architecture}

\author{\IEEEauthorblockN{Hai Dinh-Tuan}
\IEEEauthorblockA{Technische Universit\"{a}t Berlin\\
Berlin, Germany\\
hai.dinhtuan@tu-berlin.de}}

\maketitle

\begin{abstract}
Open-source 5G core implementations deploy network functions as always-on processes that consume resources even when idle. This inefficiency is most problematic in private and edge deployments with sporadic traffic. \textit{Serverless5GC} is an architecture that maps each 3GPP control-plane procedure to an independent Function-as-a-Service invocation, allowing scale-to-zero operation without modifying the standard N2 interface. The architecture decomposes 12~network functions (3GPP Release~15--17) into 31~serverless procedures, fronted by an SCTP/NGAP proxy that bridges unmodified RAN equipment to an HTTP-based serverless backend. Evaluation against Open5GS and free5GC across five traffic scenarios (idle to 50~registrations/s burst) shows that Serverless5GC achieves median registration latency of 406--522\,ms, on par with the C-based Open5GS baseline (403--606\,ms), while maintaining 100\% success across 3{,}000~registrations. A resource-time cost model shows that the serverless deployment (0.002~GB-seconds per registration) is cheaper than the always-on baseline when the cluster operates below a 0.65 duty cycle, when two or more tenants share the platform, or on managed FaaS platforms up to 595~reg/s. Under worst-case cold-start conditions where all 31~function pods are evicted simultaneously, the system sustains zero failures and converges to warm-start latency within 4--5~seconds.
\end{abstract}


\section{Introduction}
\label{sec:introduction}

The service-based architecture of the 3GPP 5G Core (5GC) consists of interconnected Network Functions (NFs). Existing open-source implementations, including Open5GS~\cite{open5gs} and free5GC~\cite{free5gc}, typically deploy these NFs as persistent, long-running processes~\cite{dinhtuan2020frameworks}. Because they draw on CPU and memory resources even when inactive, this deployment model introduces substantial inefficiencies. Such resource waste is especially noticeable in private 5G networks, enterprise environments, and emerging markets, where network traffic tends to be intermittent and idle periods are prevalent.

Serverless computing, or Function-as-a-Service (FaaS), offers an alternative: code executes only in response to events, scaling to zero when idle. This model fits the event-driven nature of 3GPP signaling procedures, where operations such as registration, authentication, and session establishment follow well-defined request-response patterns. Recent work on procedure-based decomposition~\cite{pp5gs-2023} has shown that reorganizing NFs around control plane procedures reduces inter-NF traffic and simplifies orchestration. However, it retains an always-on deployment model, leaving the idle-cost inefficiency unaddressed.

To bridge this gap, this paper proposes \textit{Serverless5GC}\footnote{The Serverless5GC implementation is publicly available at \url{https://github.com/haidinhtuan/serverless5gc}}, which maps procedure-based decomposition onto a serverless execution model. In this architecture, each 3GPP procedure becomes an independent FaaS invocation that scales to zero when idle. The well-defined request-response semantics of 3GPP signaling align directly with the FaaS execution model, and the resulting cold-start latency remains manageable within 3GPP Non-Access Stratum (NAS) timer budgets. The contributions are:
\begin{itemize}
    \item A demonstration that procedure-based 5G core decomposition maps to a serverless execution model, with 31 functions across 12 NFs (including Release~17 NFs) that achieve latency parity with always-on baselines and scale to zero when idle.
    \item An SCTP/NGAP proxy that bridges the standard N2 interface to the serverless HTTP backend, compatible with unmodified RAN (Radio Access Network) equipment and simulators.
    \item A comparative evaluation against Open5GS and free5GC under five traffic scenarios, quantifying end-to-end latency, cold-start behavior, and resource consumption.
    \item A resource-time cost model that derives the break-even conditions under which serverless 5G core deployment is cheaper than always-on alternatives across four deployment scenarios (self-hosted, cluster shutdown, multi-tenant, managed FaaS).
\end{itemize}

\section{Background and Related Work}
\label{sec:background}

\subsection{5G Core Service-Based Architecture}

The 3GPP 5G System (5GS) architecture~\cite{3gpp-23501} defines NFs communicating via HTTP/2 service-based interfaces (SBI). Core NFs include the Access and Mobility Management Function (AMF), Session Management Function (SMF), Unified Data Management (UDM), Unified Data Repository (UDR), Authentication Server Function (AUSF), Network Repository Function (NRF), Policy Control Function (PCF), and Network Slice Selection Function (NSSF). The implementation also incorporates Release~17-compliant versions of the Network Data Analytics Function (NWDAF, TS~29.520), Charging Function (CHF, TS~32.291), Network Slice Admission Control Function (NSACF, TS~29.536), and Binding Support Function (BSF, TS~29.521). Control plane procedures follow standardized flows in TS~23.502~\cite{3gpp-23502}. Open-source implementations include Open5GS~\cite{open5gs} (11 NFs, C) and free5GC~\cite{free5gc} (14 NFs, Go), both deploying NFs as persistent containerized processes. Benchmarking studies~\cite{oran-bench-2024, 5gc-bench-2023} have established baseline performance expectations for these implementations.

\subsection{Serverless Computing for Network Functions}

Prior work on serverless networking has focused on the \textit{platform} level rather than 5G core decomposition. Serverless Network Functions (SNF)~\cite{snf-2020} proposed an NFaaS platform with flowlet-granularity work assignment for network function chains; QFaaS~\cite{qfaas-2022} reduced function chain latency by 40\% using QUIC on OpenFaaS (open-source FaaS platform); and DirectFaaS~\cite{directfaas-2024} separated control/data flows in serverless chains, cutting execution time by 30.9\%. These platform-level optimizations are complementary to our application-level decomposition.

For 5G core networks specifically, Watanabe et al.~\cite{proc5gc-2023} introduced Proc5GC, a per-UE reactive stateless model using procedure-based decomposition on Linux VMs. Akashi et al.~\cite{cloud5gc-2024} extended this into Cloud5GC, deploying a Registration-only implementation on AWS (Lambda, DynamoDB, SQS). However, Cloud5GC is tightly coupled to proprietary AWS services, unsuitable for on-premises deployments, and delegates SCTP termination to a separate on-premises gateway rather than providing native N2 support.

\begin{table}[h!]
\centering
\caption{Comparison with Related 5GC Re-architectures}
\label{tab:related-work}
\resizebox{\columnwidth}{!}{%
\begin{tabular}{@{}>{\raggedright\arraybackslash}p{1.6cm}
                >{\raggedright\arraybackslash}p{1.7cm}
                c
                >{\centering\arraybackslash}p{0.8cm}
                >{\raggedright\arraybackslash}p{1.5cm}@{}}
\toprule
\textbf{Proposal} & \textbf{Platform} & \textbf{NFs/Fn.} & \textbf{N2} & \textbf{Focus} \\
\midrule
PP5GS~\cite{pp5gs-2023} & Kubernetes (always-on) & 8/4 & Sim. & Latency \\
Proc5GC~\cite{proc5gc-2023} & Linux virtual machines & --/1 & Yes & Procedure decomposition \\
Cloud5GC~\cite{cloud5gc-2024} & AWS Lambda & --/1 & No & Feasibility \\
SNF~\cite{snf-2020} & Custom & --/-- & No & Data-plane \\
\textbf{Serverless5GC} & \textbf{OpenFaaS/K3s} & \textbf{12/31} & \textbf{Yes} & \textbf{Cost efficiency} \\
\bottomrule
\end{tabular}}
\end{table}

\subsection{Procedure-Based Decomposition}

Goshi et al.\ introduced procedure-based NF decomposition in PP5GS~\cite{pp5gs-2023}, achieving up to 34\% fewer computing resources versus stateful and up to 55\% versus stateless baselines, with at least 40\% signaling traffic reduction. Subsequent work proposed piggyback state management by appending UE context to inter-NF HTTP messages~\cite{pp5gs-globecom-2023}, reducing procedure completion time by up to 44\% (Registration) and 70\% (Deregistration).

The present work addresses a different problem: \textit{cost efficiency under variable load}. PP5GS optimizes \textit{performance} while retaining always-on deployments; Serverless5GC optimizes \textit{cost} by mapping procedure-based decomposition onto a serverless model where functions scale to zero. This distinction is critical for private networks and edge deployments where traffic is sporadic.

In terms of scope, Cloud5GC covers only the Registration procedure on proprietary infrastructure, delegates SCTP to an on-premises gateway, and provides no baseline comparisons. Serverless5GC provides: (1)~a vendor-independent implementation with 31 functions across 12 NFs on open-source infrastructure (OpenFaaS/K3s); (2)~an SCTP/NGAP proxy for standard N2 connectivity; (3)~Release~17 NF coverage including NSACF; and (4)~a comparative evaluation against Open5GS and free5GC. Table~\ref{tab:related-work} summarizes these distinctions.


\section{System Architecture}
\label{sec:architecture}

\subsection{Procedure-as-a-Function Design}

Building on the procedure-based decomposition principle established by PP5GS~\cite{pp5gs-2023}, the Serverless5GC architecture maps each 3GPP procedure (as defined in TS~23.502~\cite{3gpp-23502}) to an individual serverless function. Where PP5GS consolidates the procedure-specific logic of multiple NFs into a single always-on Per-Procedure NF (PPNF) to reduce inter-NF signaling traffic and procedure completion time, the present approach takes the opposite direction: each procedure becomes an independent FaaS invocation that is instantiated on demand and scales to zero when idle. Table~\ref{tab:functions} lists the complete mapping of functions to NFs and procedures. Each function is implemented in Go using the OpenFaaS \texttt{go-http} template, exposing a single \texttt{Handle()} entry point that receives an HTTP request and returns a response.

\begin{table}[ht]
\centering
\caption{Procedure-as-a-Function Decomposition}
\label{tab:functions}
\resizebox{\columnwidth}{!}{%
\begin{tabular}{@{}llc@{}}
\toprule
\textbf{NF} & \textbf{Procedure} & \textbf{Count} \\
\midrule
AMF & Registration, Deregistration, Service Request, & \multirow{2}{*}{6} \\
    & PDU Session Relay, Handover, Auth Initiate & \\
\midrule
SMF & PDU Create, Update, Release, N4 Setup & 4 \\
\midrule
NRF & Register, Discover, Status Notify & 3 \\
\midrule
CHF\textsuperscript{\dag} & Charging Create, Update, Release & 3 \\
\midrule
BSF\textsuperscript{\dag} & Binding Register, Discover, Deregister & 3 \\
\midrule
UDM & Generate Auth Data, Get Subscriber Data & 2 \\
\midrule
UDR & Data Read, Data Write & 2 \\
\midrule
PCF & Policy Create, Policy Get & 2 \\
\midrule
NSACF\textsuperscript{\dag} & Slice Availability Check, Update Counters & 2 \\
\midrule
NWDAF\textsuperscript{\dag} & Analytics Subscribe, Data Collect & 2 \\
\midrule
AUSF & Authenticate & 1 \\
\midrule
NSSF & Slice Select & 1 \\
\midrule
\multicolumn{2}{r}{\textbf{Total}} & \textbf{31} \\
\bottomrule
\multicolumn{3}{l}{\scriptsize \textsuperscript{\dag}Feature-gated NF}
\end{tabular}}
\end{table}


The Release~17-compliant NFs (NWDAF, CHF, NSACF, BSF) are feature-gated via environment variables, so the system can be evaluated with and without R17 overhead. These NFs are well suited to serverless deployment because their invocation patterns are event-driven and bursty (analytics collection, charging lifecycle, stateless admission checks, CRUD bindings).

Inter-function communication follows the 3GPP SBI pattern via the OpenFaaS gateway using HTTP. During registration, the AMF calls AUSF for authentication, which in turn retrieves authentication vectors from UDM; with R17 enabled, the AMF additionally calls NSACF for slice admission control, and the SMF calls PCF, BSF, CHF, and NSACF during PDU session establishment. The gateway routes calls to available replicas, creating new instances as needed.


\subsection{Decomposition Methodology}

The decomposition maps each distinct service operation in the 3GPP SBI specifications (e.g., \texttt{Namf\_Communication}, \texttt{Nsmf\_PDUSession}) to a standalone serverless function whenever the operation corresponds to a complete request-response transaction. The granularity is guided by three principles: (1)~\textit{isolation}: one procedure type per function; (2)~\textit{statelessness}: UE and session state externalized to Redis/etcd; and (3)~\textit{minimal fan-out}: orchestrator functions call downstream functions sequentially.

\subsection{State Management}

Serverless functions are inherently stateless, but 5G core procedures require persistent state~\cite{dinhtuan2022ms2m}: User Equipment (UE) context, Protocol Data Unit (PDU) session data, and authentication vectors must survive across function invocations. The system employs a dual-store architecture fulfilling the role of the 3GPP Unstructured Data Storage Function (UDSF, TS~29.598):

\begin{itemize}
    \item \textit{Redis (transient UDSF)}: Stores UE context (registration state, security context), PDU session data, authentication vectors, charging sessions, slice admission counters, PCF bindings, and analytics subscriptions. Keys follow the pattern \texttt{ue:\{supi\}}, \texttt{pdu-sessions/\{session-id\}}, \texttt{auth-vectors/\{supi\}}, \texttt{charging-sessions/\{id\}}, \texttt{nsacf-counters/\{sst\}-\{sd\}}, and \texttt{bsf-bindings/\{id\}}.
    \item \textit{etcd (NRF UDSF)}: Serves as the NRF service registry, storing NF instance profiles and supporting service discovery through key-prefix queries. Using etcd rather than a dedicated NRF database decouples discovery from NF lifecycle and provides strong consistency guarantees.
\end{itemize}

All functions access state through a common \texttt{KVStore} interface that abstracts the underlying store and allows mock injection for testing. This design is functionally equivalent to a UDSF~\cite{udsf-patent-2019,3gpp-29598}, where unstructured data associated with UE and NF instances is stored externally so that NF instances remain stateless.


\textit{Concurrency.} Data races are mitigated through Redis's single-threaded execution, isolated key schemas, and \texttt{WATCH}/\texttt{MULTI}/\texttt{EXEC} transactions. In practice, the SCTP proxy serializes the NAS state machine per UE, preventing concurrent writes to the same key.


\subsection{SCTP/NGAP Proxy}

The N2 interface between the next-generation Node B (gNB) and the AMF uses Stream Control Transmission Protocol (SCTP) transport with NG Application Protocol (NGAP, TS~38.413~\cite{3gpp-38413}) encoding. Since serverless functions are invoked via HTTP, a protocol translation layer is required.

\begin{figure*}[t]
\centering
\includegraphics[width=\textwidth]{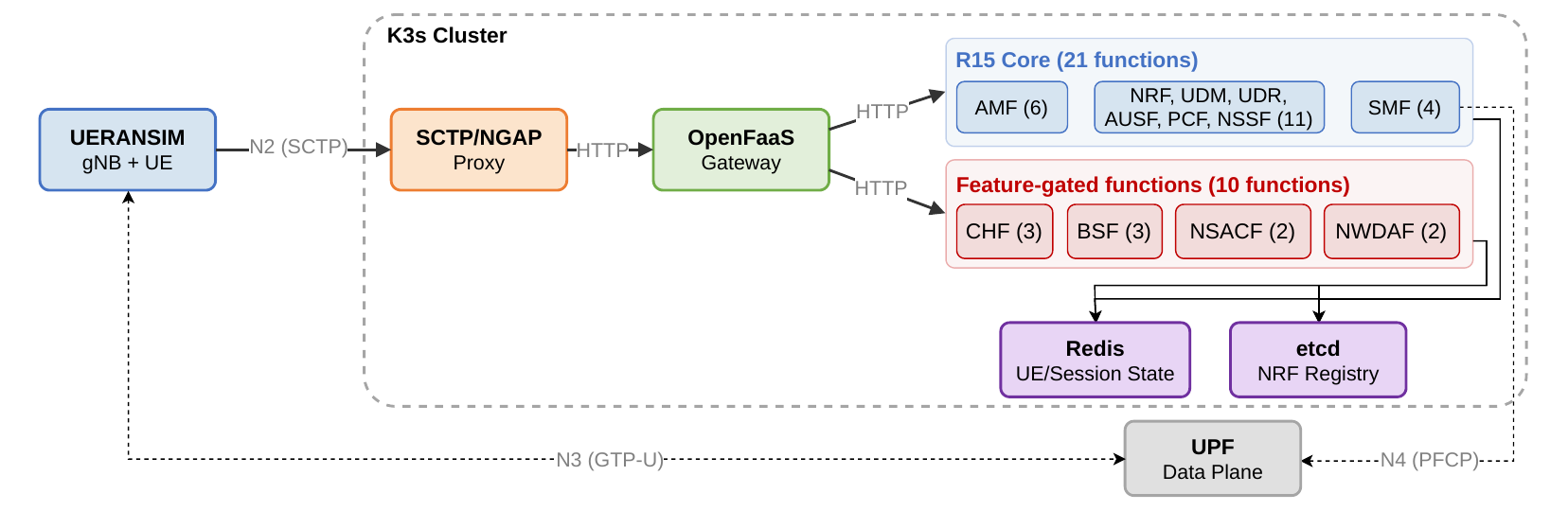}
\caption{Serverless 5GC architecture. The SCTP/NGAP proxy bridges the N2 interface (SCTP) to HTTP-based function invocations. R15 core functions (21) handle standard procedures; 10 feature-gated functions add analytics, charging, admission control, and PCF binding. All functions share Redis for state and etcd for NRF service discovery. UPF (User Plane Function, out of scope) is connected via N4 (PFCP) from SMF and N3 (GTP-U) from the gNB for completeness.}
\label{fig:architecture}
\end{figure*}

The SCTP proxy (Fig.~\ref{fig:architecture}) accepts SCTP connections from gNBs, decodes NGAP and NAS~\cite{3gpp-24501} messages, maintains per-UE state machines for signaling flow tracking, and converts HTTP/JSON responses from serverless functions back into security-protected NGAP/NAS messages.


The proxy routes calls to the OpenFaaS gateway via a \texttt{CoreBackend} interface using Go's \texttt{http.Client} with persistent keep-alive connections. This eliminates per-invocation TCP (Transmission Control Protocol) handshake overhead and accounts for the observed latency parity with direct-call architectures.

\textit{Transport limitations.} Terminating SCTP at the proxy sacrifices two SCTP properties: (1)~\textit{multi-homing} path redundancy, which must be compensated by redundant proxy replicas behind an SCTP-aware load balancer; and (2)~\textit{head-of-line (HoL) free streaming}, mitigated by Go's keep-alive TCP connection pooling and the sub-millisecond intra-cluster RTT. The \texttt{CoreBackend} interface accommodates future transport upgrades (e.g., HTTP/3).


\subsection{Deployment Infrastructure}

The system deploys on a single-node K3s (lightweight Kubernetes) cluster with OpenFaaS (function lifecycle and scaling), Redis (256\,MB, LRU eviction), etcd (NRF registry), and the SCTP proxy on NGAP port 38412. Each function is packaged as a multi-stage Docker container producing a static Go binary on Alpine Linux, served via the OpenFaaS \texttt{go-http} template SDK.

As shown in Table~\ref{tab:functions}, the system decomposes 12 NFs into 31 serverless functions, comparable in coverage to Open5GS (11 NFs) and free5GC (14 NFs), with each procedure scaling independently.

\section{Evaluation}
\label{sec:evaluation}

\subsection{Experimental Setup}

The evaluation deploys three system configurations on cloud virtual machines in the same region:

\begin{enumerate}
    \item \textit{Serverless5GC}: The Procedure-as-a-Function architecture on a 4-vCPU/8\,GB VM running K3s~v1.31, OpenFaaS~CE, Redis~7, etcd~3.5.17, and the SCTP/NGAP proxy (Go~1.25). Internal per-function execution times are measured via the OpenFaaS gateway metrics.
    \item \textit{Open5GS}~\cite{open5gs}: v2.7.2 (C-based), deployed via Docker Compose on a 4-vCPU/8\,GB VM.
    \item \textit{free5GC}~\cite{free5gc}: v4.2.0 (Go-based), deployed via Docker Compose on a 4-vCPU/8\,GB VM.
\end{enumerate}

A separate 4-vCPU/8\,GB load generator VM hosts UERANSIM (5G UE/RAN simulator)~v3.2.6~\cite{ueransim} and the traffic generation scripts. A dedicated monitoring VM of identical specification runs Prometheus~v2.51 (5-second scrape interval) and node exporters, collecting metrics from all target VMs and the load generator. All VMs run Ubuntu~24.04.

All VMs reside within the same virtual data center, connected via a private LAN (Local Area Network) with sub-millisecond inter-VM latency. Tests are executed sequentially (one target at a time) to prevent cross-system interference.

\subsection{Traffic Scenarios}

The evaluation covers five traffic scenarios, each representing a different operational regime as shown in Table \ref{tab:scenarios}:

\begin{table}[ht]
\centering
\caption{Traffic Scenarios}
\label{tab:scenarios}
\footnotesize
\resizebox{\columnwidth}{!}{%
\begin{tabular}{@{}lrllrl@{}}
\toprule
\textbf{Scenario} & \textbf{UEs} & \textbf{Rate} & \textbf{Duration} & \textbf{PDU/UE} & \textbf{Description} \\
\midrule
Idle   & 0    & --       & 10\,min & -- & No traffic (scale-to-zero) \\
Low    & 100  & 1/s      & 10\,min & 1  & Sporadic registrations \\
Medium & 500  & 5/s      & 10\,min & 2  & Moderate steady state \\
High   & 1000 & 20/s     & 10\,min & 3  & Sustained load \\
Burst  & 500  & 50/s     & 5\,min  & 2  & Peak overload \\
\bottomrule
\end{tabular}%
}
\end{table}

Each scenario is repeated three times per target configuration. CPU utilization, memory consumption, function invocation counts, and response latencies are collected via Prometheus. The evaluation exclusively targets \textit{private and enterprise 5G networks} (campus deployments, factory floors, remote-area cells) where 100--1{,}000 UEs represent realistic operational loads. Carrier-grade networks serving tens of thousands UEs are outside the scope of this work.

\textit{UE arrival model:} UERANSIM processes UE state machines sequentially within a single process, limiting the effective registration throughput per instance to 50~UEs/s~\cite{ueransim}. To reach the target UE counts (500--1{,}000), multiple UERANSIM instances are launched in batches of 100~UEs, staggered by 30~seconds. In the low scenario (100~UEs), a single batch completes within 2\,s. In medium, high, and burst scenarios, only the first batch (10--20\% of UEs) arrives within the first 2\,s; the remaining 80--90\% arrive 30--270\,s later (medium/burst: up to 120\,s; high: up to 270\,s). This batching structure affects the interpretation of cold-start measurements (Section~\ref{sec:coldstart}).

Table~\ref{tab:latency-comparison} shows end-to-end registration latency. Serverless5GC achieves median latency of 406--522\,ms across scenarios, matching the C-based Open5GS baseline (403--606\,ms), while free5GC is 6--8$\times$ slower at the median.

\begin{table}[ht]
\centering
\caption{End-to-End Registration Latency (ms)}
\label{tab:latency-comparison}
\footnotesize
\resizebox{\columnwidth}{!}{%
\begin{tabular}{@{}l rrr rrr rrr@{}}
\toprule
 & \multicolumn{3}{c}{\textbf{Serverless5GC}} & \multicolumn{3}{c}{\textbf{Open5GS}} & \multicolumn{3}{c}{\textbf{free5GC}} \\
\cmidrule(lr){2-4} \cmidrule(lr){5-7} \cmidrule(lr){8-10}
\textbf{Scenario} & \textbf{p50} & \textbf{p95} & \textbf{p99} & \textbf{p50} & \textbf{p95} & \textbf{p99} & \textbf{p50} & \textbf{p95} & \textbf{p99} \\
\midrule
Low    & 414  & 686   & 698  & 406  & 500   & 1{,}453  & 3{,}234  & 4{,}466  & 5{,}196  \\
Medium & 406  & 3{,}770  & 5{,}039  & 410  & 3{,}863  & 5{,}170  & 3{,}257  & 5{,}847  & 6{,}875  \\
High   & 522  & 9{,}147  & 9{,}915  & 606  & 8{,}666  & 10{,}185 & 3{,}592  & 10{,}878 & 12{,}661 \\
Burst  & 435  & 3{,}762  & 5{,}083  & 403  & 4{,}122  & 5{,}166  & 3{,}008  & 5{,}775  & 6{,}760  \\
\bottomrule
\end{tabular}%
}
\end{table}

\begin{table}[ht]
\centering
\caption{PDU Session Establishment Latency (ms)}
\label{tab:pdu-latency}
\footnotesize
\resizebox{\columnwidth}{!}{%
\begin{tabular}{@{}l rrr rrr rrr@{}}
\toprule
 & \multicolumn{3}{c}{\textbf{Serverless5GC}} & \multicolumn{3}{c}{\textbf{Open5GS}} & \multicolumn{3}{c}{\textbf{free5GC}} \\
\cmidrule(lr){2-4} \cmidrule(lr){5-7} \cmidrule(lr){8-10}
\textbf{Scenario} & \textbf{p50} & \textbf{p95} & \textbf{p99} & \textbf{p50} & \textbf{p95} & \textbf{p99} & \textbf{p50} & \textbf{p95} & \textbf{p99} \\
\midrule
Low    & 195  & 363    & 425     & 109  & 157    & 212     & 2{,}680  & 3{,}064  & 3{,}161  \\
Medium & 169  & 348    & 16{,}628 & 113  & 357    & 16{,}644 & 1{,}919  & 2{,}849  & 16{,}164 \\
High   & 186  & 423    & 16{,}707 & 146  & 361    & 16{,}970 & 1{,}532  & 3{,}075  & 17{,}048 \\
Burst  & 163  & 382    & 16{,}619 & 119  & 339    & 16{,}730 & 1{,}810  & 2{,}817  & 16{,}591 \\
\bottomrule
\end{tabular}%
}
\end{table}

\subsection{PDU Session Latency}

Table~\ref{tab:pdu-latency} shows PDU session establishment latency. The serverless architecture achieves median PDU latency of 163--195\,ms, comparable to Open5GS (109--146\,ms) and 8--14$\times$ faster than free5GC (1{,}532--2{,}680\,ms). The increased p99 values (average 16{,}678\,ms across all three systems) observed under medium, high, and burst loads reflect UERANSIM's sequential UE state machine processing: the last UEs in each batch experience queuing delays at the load generator, not at the core network. This artifact affects all systems equally and does not indicate a system-specific bottleneck.

\textit{Success rates.} The serverless architecture achieves 100\% registration and PDU session success across all scenarios (6{,}300/6{,}300 total). Open5GS drops 4~registrations under high load (99.94\% overall) and free5GC drops 1 (99.98\%). Results are reproducible across runs: cross-run $\sigma$ of the median registration latency ranges from 3 to 47\,ms for Serverless5GC and 13 to 97\,ms for Open5GS, while free5GC exhibits higher run-to-run variability (54 to 564\,ms).

Table~\ref{tab:function-latency} shows how long each function takes to execute (HTTP handler only, excluding SCTP and network overhead). Each UE registration triggers 8 function invocations in total; summed, they complete in 16.49\,ms, corresponding to just 3.7\% of the 444\,ms median end-to-end latency. The remaining 96.3\% is the 3GPP protocol itself: NAS round trips over the radio and N2 interfaces (Registration Request, Authentication Request/Response, Security Mode Command/Complete, and Registration Accept/Complete), plus UE-side cryptographic processing of the authentication challenge.

\begin{table}[ht]
\centering
\caption{Function Execution Time, High Scenario (ms)}
\label{tab:function-latency}
\footnotesize
\resizebox{\columnwidth}{!}{%
\begin{tabular}{@{}lrr@{}}
\toprule
\textbf{Function} & \textbf{Avg (ms)} & \textbf{Invocations} \\
\midrule
\texttt{amf-initial-registration} & 7.68 & 1{,}000 \\
\texttt{amf-auth-initiate}        & 2.33 & 2{,}000 \\
\texttt{udm-generate-auth-data}   & 0.84 & 2{,}000 \\
\texttt{smf-pdu-session-create}   & 0.93 & 1{,}000 \\
\texttt{ausf-authenticate}        & 0.80 & 1{,}000 \\
\texttt{udm-get-subscriber-data}  & 0.74 & 1{,}000 \\
\bottomrule
\end{tabular}%
}
\par\vspace{1pt}
{\scriptsize Excludes proxy/network overhead.}
\end{table}

Table~\ref{tab:function-scaling} shows that per-function execution times are \emph{virtually independent of load level}: even at 10$\times$ the invocation rate (low to high), the maximum deviation for any function is under 5\%. This confirms that the stateless, externalized-state design eliminates contention between concurrent invocations. Each UE registration triggers exactly 8~function invocations across all scenarios, validating the deterministic call graph of the Procedure-as-a-Function model.

\begin{table}[ht]
\centering
\caption{Function Execution Time Across Load Levels (ms)}
\label{tab:function-scaling}
\footnotesize
\resizebox{\columnwidth}{!}{%
\begin{tabular}{@{}lrrrr@{}}
\toprule
\textbf{Function} & \textbf{Low} & \textbf{Medium} & \textbf{High} & \textbf{Burst} \\
\midrule
\texttt{amf-initial-reg.} & 7.58 & 7.56 & 7.68 & 7.76 \\
\texttt{amf-auth-initiate}  & 2.29 & 2.28 & 2.33 & 2.33 \\
\texttt{udm-gen-auth-data}  & 0.82 & 0.82 & 0.84 & 0.85 \\
\texttt{smf-pdu-create}     & 0.91 & 0.91 & 0.93 & 0.92 \\
\texttt{ausf-authenticate}  & 0.78 & 0.80 & 0.80 & 0.81 \\
\texttt{udm-get-sub-data}   & 0.73 & 0.72 & 0.74 & 0.75 \\
\bottomrule
\multicolumn{5}{l}{\scriptsize Internal execution times only (excludes proxy/SCTP/network overhead).}
\end{tabular}%
}
\end{table}


The latency measurements above reflect warm-start conditions. Cold-start behavior is evaluated explicitly in the following subsection.

\subsection{Cold-Start Storm Behavior}
\label{sec:coldstart}

A key concern for serverless 5G core deployments is the \emph{cold-start storm} scenario: after a prolonged idle period, the autoscaler has scaled all functions to zero replicas, and a burst of UEs simultaneously attempts registration, forcing the container runtime to provision all function pods concurrently while signaling requests are already in flight. To quantify this worst case, a dedicated cold-start experiment is conducted using the same traffic scenarios as the warm-start evaluation (Table~\ref{tab:scenarios}).

\textit{Methodology.} For each run, all 31 function pods are force-deleted (excluding infrastructure: Redis, etcd) and UERANSIM UE registrations are started \emph{simultaneously}, ensuring the first UEs hit the system while function containers are still initializing. The Kubernetes deployment controller immediately recreates the deleted pods, but each function must complete container startup and Go runtime initialization before serving requests. Each scenario is run 3~times and averages are reported.

Due to the batched UE arrival model (Section~\ref{sec:evaluation}), the first 100-UE batch arrives within 2\,s, overlapping with the 4\,s pod recreation window. In the low scenario, every UE is affected. In medium, high, and burst scenarios, only the first batch (10--20\%) overlaps with cold-start; the remaining 80--90\% hit warm replicas.

\begin{table}[ht]
\centering
\caption{Cold-Start vs.\ Warm-Start Registration Latency (ms)}
\label{tab:coldstart-comparison}
\footnotesize
\resizebox{\columnwidth}{!}{%
\begin{tabular}{@{}l rrr rrr r@{}}
\toprule
 & \multicolumn{3}{c}{\textbf{Cold-Start}} & \multicolumn{3}{c}{\textbf{Warm-Start}} & \\
\cmidrule(lr){2-4} \cmidrule(lr){5-7}
\textbf{Scenario} & \textbf{p50} & \textbf{p95} & \textbf{p99} & \textbf{p50} & \textbf{p95} & \textbf{p99} & \textbf{$\Delta$ p99} \\
\midrule
Low    & 5{,}037 & 5{,}791 & 5{,}938 & 414  & 686  & 698  & $+$5{,}240\,ms \\
Medium & 366     & 5{,}340 & 5{,}729 & 406  & 3{,}770 & 5{,}039 & $+$690\,ms \\
High   & 377     & 8{,}995 & 10{,}873 & 522  & 9{,}147 & 9{,}915 & $+$958\,ms \\
Burst  & 361     & 6{,}059 & 6{,}485 & 435  & 3{,}762 & 5{,}083 & $+$1{,}402\,ms \\
\bottomrule
\end{tabular}%
}
\end{table}

\begin{figure}[t]
    \centering
    \begin{tikzpicture}
    \begin{axis}[
        ybar,
        bar width=8pt,
        width=0.88\columnwidth,
        height=6cm,
        ylabel={Registration Latency, p99 (ms)},
        symbolic x coords={Low, Medium, High, Burst},
        xtick=data,
        nodes near coords,
        nodes near coords align={vertical},
        every node near coord/.append style={font=\tiny, /pgf/number format/.cd, fixed, precision=0},
        ymin=0, ymax=12500,
        ytick={0,5000,10000},
        scaled y ticks=false,
        legend style={at={(0.5,-0.20)}, anchor=north, legend columns=2, font=\small},
        enlarge x limits=0.2,
        ymajorgrids=true,
    ]
    \addplot+[error bars/.cd, y dir=both, y explicit] coordinates { (Low,5938) +- (0,473) (Medium,5729) +- (0,816) (High,10873) +- (0,312) (Burst,6485) +- (0,171) };
    \addlegendentry{cold}
    \addplot+[error bars/.cd, y dir=both, y explicit] coordinates { (Low,698) +- (0,10) (Medium,5039) +- (0,161) (High,9915) +- (0,171) (Burst,5083) +- (0,158) };
    \addlegendentry{warm}
    \end{axis}
\end{tikzpicture}
    \caption{p99 registration latency under cold-start storm vs.\ warm-start conditions (3-run averages, whiskers show $\pm\sigma$). The low scenario (100~UEs) shows an 8.5$\times$ tail penalty as all UEs hit initializing containers; in larger sets, the cold-start penalty is visible only in the tail ($+$690--$+$1{,}402\,ms at p99).}
    \label{fig:coldstart-latency}
\end{figure}

Table~\ref{tab:coldstart-comparison} presents the results. The critical metric is p99: it captures the tail where UEs arrive while pods are still initializing, and it must remain below the 15\,s NAS T3510 retransmission timer. In the low scenario (100~UEs, single batch), all UEs arrive within the 4\,s pod recreation window; p99 reaches 5{,}938\,ms, consuming 39.6\% of the T3510 budget. In multi-batch scenarios (500--1{,}000~UEs), only the first 100-UE batch overlaps with cold-start; $\Delta$\,p99 ranges from $+$690\,ms (medium) to $+$1{,}402\,ms (burst), peaking at 10{,}873\,ms in the high scenario (72.5\% of budget).

Across all 12 cold-start runs, the system achieves \textit{100\% registration success} without T3510 timer expirations. The system converges to warm-start performance within 4--5\,seconds, possible by the small memory footprint of each function (approximately 15\,MB) and local container image caching.


\subsection{Resource Consumption}

\begin{figure}[t]
    \centering
    \begin{tikzpicture}
    \begin{axis}[
        ybar,
        bar width=11pt,
        width=0.88\columnwidth,
        height=6cm,
        ylabel={CPU Rate (s\,min$^{-1}$)},
        symbolic x coords={Low, Medium, High, Burst},
        xtick=data,
        nodes near coords,
        nodes near coords align={vertical},
        every node near coord/.append style={font=\tiny, /pgf/number format/.cd, fixed, precision=1},
        ymin=0, ymax=30,
        legend style={at={(0.5,-0.20)}, anchor=north, legend columns=3, font=\small},
        enlarge x limits=0.3,
        ymajorgrids=true,
    ]
    \addplot coordinates { (Low,7.9) (Medium,10.0) (High,13.2) (Burst,12.5) };
    \addlegendentry{Serverless5GC}
    \addplot coordinates { (Low,2.3) (Medium,4.4) (High,6.9) (Burst,6.9) };
    \addlegendentry{Open5GS}
    \addplot coordinates { (Low,5.6) (Medium,20.0) (High,25.5) (Burst,24.2) };
    \addlegendentry{Free5GC}
    \end{axis}
\end{tikzpicture}
    \caption{CPU resource consumption rate (seconds per minute of measurement window) across traffic scenarios. Burst uses a 5-minute window; all other scenarios use a 10-minute window.}
    \label{fig:resources}
\end{figure}

Table~\ref{tab:resource-comparison} compares the resource footprint of the three systems. Data represents the average of 3 runs. CPU and memory are measured at the node level via Prometheus \texttt{node\_cpu\_seconds\_total} and \texttt{node\_memory\_MemAvailable\_bytes}, filtered by VM label to isolate each target system. CPU values are reported as \textit{busy CPU seconds per minute} (sum of non-idle CPU deltas over the measurement window, divided by window length: 10\,min for steady-state, 5\,min for burst). Memory is reported as \textit{used memory} (total minus available) in MB.

The serverless architecture has a higher baseline CPU (7.6\,s\,min$^{-1}$ idle vs.\ 2.0 for Open5GS, 3.4 for free5GC) due to the K3s control plane, OpenFaaS, Redis, and etcd. However, CPU growth from idle to high load is only $+$74\% (7.6$\to$13.2), compared to $+$245\% for Open5GS (2.0$\to$6.9) and $+$650\% for free5GC (3.4$\to$25.5). Memory consumption is stable for the serverless architecture (2{,}105--2{,}138\,MB, $+$1.6\% from idle to high), as function replicas carry no per-UE state. The baselines show more variation (Open5GS $+$17\%, free5GC $+$15\%). Peak utilization reaches 10.6\% of the 4-vCPU capacity (free5GC, high scenario).

%

The higher serverless baseline is attributable to the orchestration platform (K3s + OpenFaaS account for 1{,}130\,MB, or 53.7\% of the 2{,}105\,MB total), not the function workload ($31 \times 15$\,MB $= 465$\,MB, comparable to a single Open5GS NF process).

\begin{table}[ht]
\centering
\caption{Node-Level Resource Consumption}
\label{tab:resource-comparison}
\footnotesize
\resizebox{\columnwidth}{!}{%
\begin{tabular}{@{}l rr rr rr@{}}
\toprule
 & \multicolumn{2}{c}{\textbf{Serverless5GC}} & \multicolumn{2}{c}{\textbf{Open5GS}} & \multicolumn{2}{c}{\textbf{free5GC}} \\
\cmidrule(lr){2-3} \cmidrule(lr){4-5} \cmidrule(lr){6-7}
\textbf{Scenario} & \textbf{CPU} & \textbf{Mem} & \textbf{CPU} & \textbf{Mem} & \textbf{CPU} & \textbf{Mem} \\
\midrule
Idle   & 7.6 & 2105 & 2.0 & 1368 & 3.4 & 1378 \\
Low    & 7.9 & 2130 & 2.3 & 1442 & 5.6 & 1463 \\
Medium & 10.0 & 2136 & 4.4 & 1524 & 20.0 & 1522 \\
High   & 13.2 & 2138 & 6.9 & 1605 & 25.5 & 1584 \\
Burst  & 12.5 & 2134 & 6.9 & 1493 & 24.2 & 1535 \\
\bottomrule
\end{tabular}%
}
\par\vspace{1pt}
{\scriptsize CPU: busy seconds\,min$^{-1}$ (non-idle delta $\div$ window). Mem: used MB (8\,GB $-$ available).}
\end{table}

\subsection{Cost Model}
\label{sec:cost-model}

The evaluation data (Tables~\ref{tab:resource-comparison} and~\ref{tab:function-latency}) permit a formal cost comparison between the serverless and always-on deployment models using the platform-independent resource-time metric (GB-seconds). The key measured quantities are:

\begin{itemize}
    \item Serverless idle memory $M_s = 2{,}105$\,MB (Table~\ref{tab:resource-comparison}), decomposed as platform base $M_p = 1{,}640$\,MB (K3s: 820\,MB, OpenFaaS: 310\,MB, Redis: 85\,MB, etcd: 65\,MB, operating system: 360\,MB) plus function replicas $M_f = 465$\,MB ($31 \times 15$\,MB).
    \item Always-on baseline $M_a = 1{,}368$\,MB: Open5GS idle memory. Open5GS is chosen as the always-on reference because it achieves latency parity with Serverless5GC (Table~\ref{tab:latency-comparison}) and has a lower idle footprint than free5GC (1,368\,MB vs.\ 1,378\,MB), making it the more conservative and fairer cost comparison.
    \item Minimum infrastructure $M_i = 150$\,MB: state stores only (Redis: 85\,MB, etcd: 65\,MB), the persistent minimum when functions run on a managed FaaS platform.
    \item Per-registration resource-time $g = 0.002$\,GB-s: the total FaaS compute consumed per UE registration, computed as $128\,\text{MB} \times 16.49\,\text{ms} = 0.002$\,GB-s, where 128\,MB is the minimum AWS Lambda memory allocation (actual function footprint is approximately 15\,MB, but Lambda bills at a 128\,MB minimum) and 16.49\,ms is the summed execution time of all 8 invocations (Table~\ref{tab:function-latency}).
\end{itemize}

The total resource-time over period $T$ (in seconds) depends on the deployment scenario. Let $\lambda$ denote the average registration rate (reg/s) and $d \in [0,1]$ the \textit{duty cycle}, defined as the fraction of $T$ during which traffic is present.

\subsubsection{Self-Hosted, Platform Always On}

Both the serverless platform and always-on baseline run on dedicated VMs. Functions scale to zero during idle, but the platform (K3s, OpenFaaS, Redis, etcd) remains active:

\begin{align}
G_s &= (M_p + d \cdot M_f) \cdot T + \lambda \cdot d \cdot T \cdot g \label{eq:gs-selfhosted} \\
G_a &= M_a \cdot T \label{eq:ga}
\end{align}

Setting $G_s = G_a$ and noting that $\lambda \cdot d \cdot g$ is negligible (at 20\,reg/s: $\lambda g = 0.04$\,GB $= 41$\,MB, about 2.5\% of $M_p = 1{,}640$\,MB):

\begin{equation}
d^* \approx \frac{M_a - M_p}{M_f} = \frac{1{,}368 - 1{,}640}{465} = -0.59 \label{eq:d-selfhosted}
\end{equation}

The negative value indicates that $M_p > M_a$: the platform overhead alone (1{,}640\,MB) exceeds the always-on baseline (1{,}368\,MB). On a dedicated single-tenant VM, the serverless deployment is never cheaper than the always-on baseline at any duty cycle. This result is a direct consequence of the K3s orchestration overhead (820\,MB) and the OpenFaaS gateway (310\,MB), which together account for 1{,}130\,MB of fixed cost absent from the Docker Compose baselines.

\subsubsection{Self-Hosted, Cluster Shutdown}

If the entire K3s cluster is shut down during idle periods (e.g., via scheduling for predictable traffic patterns), the serverless cost becomes:

\begin{equation}
G_s = M_s \cdot d \cdot T + \lambda \cdot d \cdot T \cdot g
\end{equation}

Setting $G_s = G_a$:
\begin{equation}
d^* = \frac{M_a}{M_s + \lambda \cdot g} \approx \frac{M_a}{M_s} = \frac{1{,}368}{2{,}105} = 0.65 \label{eq:d-shutdown}
\end{equation}

The serverless deployment is cheaper when the cluster runs less than 65\% of the time. For a private network operating 8\,h/day ($d = 0.33$), the resource-time cost is 51\% of the always-on baseline. For 12\,h/day ($d = 0.50$), it is 77\%.

\subsubsection{Multi-Tenant Self-Hosted}

If $K$ tenants share the K3s/OpenFaaS platform, the per-tenant platform cost is $M_p/K$, while each tenant adds its own function replicas:

\begin{equation}
G_s^{(K)} = \left(\frac{M_p}{K} + d \cdot M_f\right) \cdot T
\end{equation}

Setting $G_s^{(K)} = G_a$ with $d = 1$ (worst case, all tenants always active):
\begin{align}
K^* &= \left\lceil \frac{M_p}{M_a - M_f} \right\rceil = \left\lceil \frac{1{,}640}{1{,}368 - 465} \right\rceil \nonumber \\
    &= \left\lceil \frac{1{,}640}{903} \right\rceil = 2 \label{eq:k-multitenant}
\end{align}

With two or more tenants on a shared platform, the per-tenant serverless cost falls below the always-on baseline even at full duty cycle.

\subsubsection{Managed FaaS}

On a managed platform (e.g., AWS Lambda, Google Cloud Run), the orchestration overhead ($M_p$) is absorbed by the provider. Only the state stores ($M_i = 150$\,MB as managed services) and per-invocation costs remain:

\begin{equation}
G_s = M_i \cdot T + \lambda \cdot T \cdot g
\end{equation}

Setting $G_s = G_a$:
\begin{equation}
\lambda^* = \frac{M_a - M_i}{g} = \frac{1.189\,\text{GB}}{0.002\,\text{GB-s}} = 595\,\text{reg/s} \label{eq:lambda-managed}
\end{equation}

Below 595\,reg/s, the managed serverless deployment consumes less resource-time than the always-on baseline. 

\subsubsection{Summary}

Table~\ref{tab:cost-model} summarizes the four deployment scenarios.

\begin{table}[h]
\centering
\caption{Cost Model: Break-Even Conditions}
\label{tab:cost-model}
\footnotesize
\begin{tabular}{@{}lll@{}}
\toprule
\textbf{Scenario} & \textbf{Condition} & \textbf{Threshold} \\
\midrule
Self-hosted, platform on   & Never cheaper     & $M_p > M_a$ \\
Self-hosted, cluster off   & $d < d^*$         & $d^* = 0.65$ \\
Multi-tenant self-hosted   & $K \geq K^*$      & $K^* = 2$ \\
Managed FaaS               & $\lambda < \lambda^*$ & $\lambda^* = 595$\,reg/s \\
\bottomrule
\end{tabular}
\end{table}

This model identifies the orchestration platform overhead as the dominant cost factor for self-hosted deployments: the 1{,}130\,MB K3s/OpenFaaS footprint must be amortized across idle periods (cluster shutdown) or tenants (shared platform) to offset the always-on baseline. On managed FaaS platforms, this overhead is absent, and the per-invocation cost is low enough that serverless remains cheaper up to 595\,reg/s, well above typical private network signaling rates.

\subsection{Discussion}
\label{sec:discussion}

The evaluation results highlight several points for serverless 5G core design:

\textit{Orchestration dominance.} \texttt{amf-initial-registration} accounts for 46.5\% of total function compute time but only 12.5\% of invocations, orchestrating 7 downstream calls to 5 functions. This decomposition enables fine-grained scaling: orchestration functions can be allocated more resources without over-provisioning simple lookups where UDM and AUSF functions each complete in under 1\,ms.

\textit{Burst resilience without pre-provisioning.} Under burst traffic (50~reg/s, 2.5$\times$ the sustained high rate), the serverless architecture absorbs the spike with comparable end-to-end latency (435\,ms burst vs.\ 406\,ms medium, $+$7\%) and with internal function execution time increasing by only $+$1.7\%. No pre-scaling, capacity reservation, or manual intervention was required, which suits event-driven deployments (stadiums, festivals, disaster recovery).

\textit{Protocol translation overhead.} The total internal function execution time (16.5\,ms for 8 invocations) represents 3.7\% of the end-to-end registration latency (444\,ms). The remainder is dominated by the NAS procedure inherent to 3GPP registration (NAS round trips over the radio interface) plus UE-side cryptographic processing. Serverless invocation overhead is negligible relative to protocol-inherent latency.

\textit{Complementarity with PP5GS.} PP5GS~\cite{pp5gs-2023} consolidates NFs per procedure to reduce inter-NF HTTP traffic by 17.5--40\%. The Serverless5GC model inherits this benefit, since each function already encapsulates a complete procedure, while adding scale-to-zero capability. A combined approach could apply PP5GS's piggyback state management~\cite{pp5gs-globecom-2023} within serverless functions to further reduce external state lookups, eliminating inter-NF traffic \textit{and} idle resource costs.

\textit{Toward a hybrid architecture.} A production serverless 5G core would benefit from a \textit{hybrid} deployment model: latency-critical NFs (AMF, SMF) with persistent deployment, combined with FaaS-backed R17 NFs (NWDAF, CHF, NSACF) that are invoked sporadically and tolerate moderate cold-start latency. The Procedure-as-a-Function decomposition enables this hybrid placement directly, since each procedure is an independent deployment unit. Service affinity analysis~\cite{dinhtuan2024saga} could further inform placement decisions by co-locating frequently communicating functions.

\subsection{Limitations and Future Work}

(1)~\textit{Scale and platform}: the evaluation targets private networks (50~reg/s, 1{,}000~UEs); carrier-grade deployments would require multi-node K3s clusters, OpenFaaS gateway load balancing, Redis clustering, and redundant SCTP proxy replicas behind an SCTP-aware load balancer.
(2)~\textit{Cold-start and migration}: the 8.5$\times$ p99 tail penalty under low-UE cold-start conditions motivates predictive pre-scaling (e.g., time-of-day traffic models) and container checkpoint/restore~\cite{dinhtuan2025k8s} for warm replica migration without cold-start penalties.
(3)~\textit{Scope}: this work covers control plane only; extending to the UPF using eBPF or DPDK, and validating with physical gNBs and commercial UEs under real radio conditions, remain open.

\section{Conclusion}
\label{sec:conclusion}

This paper has presented Serverless5GC, an implementation of 31 serverless functions across 12 3GPP network functions on open-source infrastructure, with an SCTP/NGAP proxy for standard N2 interface compatibility.

The evaluation against Open5GS and free5GC leads to two findings. First, the Procedure-as-a-Function decomposition does not degrade signaling performance: median registration latency (406--522\,ms) matches the C-based Open5GS baseline (403--606\,ms), internal function execution (16.5\,ms for 8 invocations) accounts for 3.7\% of end-to-end latency, and 100\% registration success is maintained across all scenarios including cold-start storms. Second, the cost advantage of serverless 5G core deployment is conditional, not absolute. The resource-time cost model shows that on a single self-hosted VM, the platform infrastructure overhead (1{,}640\,MB, including K3s, OpenFaaS, Redis, etcd, and OS) exceeds the always-on baseline (1{,}368\,MB), making serverless more expensive at any traffic level. The cost advantage materializes under three conditions: (1)~the cluster is shut down during idle periods (break-even at 65\% duty cycle, i.e., cheaper for any deployment active less than 15.6\,h/day); (2)~two or more tenants share the platform; or (3)~functions run on a managed FaaS provider, where serverless is cheaper up to 595\,reg/s.

This result has a practical implication for private 5G deployments. The serverless model is not a universal replacement for always-on architectures, but it is cost-effective in the case of private networks: intermittent traffic, low duty cycles, and limited subscribers. A campus network active 8\,h/day ($d = 0.33$) consumes 51\% of the resource-time of an always-on equivalent, with no loss in latency or reliability.

\bibliographystyle{IEEEtran}
\bibliography{references}

\balance

\end{document}